\documentclass[runningheads]{llncs}
\usepackage{xcolor}
\usepackage[T1]{fontenc}
\usepackage{graphicx}
\usepackage{subfig}
\usepackage{amsmath}
\usepackage{cite}

\begin{document}
\title{Comparative study of random walks with one-step memory on complex networks}

\titlerunning{Random walks with one-step memory on complex networks}

\author{Miroslav Mirchev\orcidID{0000-0001-9899-2439} \and
Lasko Basnarkov\orcidID{0000-0001-5657-9304} \and
Igor Mishkovski\orcidID{0000-0003-1137-6102}}
\authorrunning{M. Mirchev et al.}

\institute{Ss. Cyril and Methodius University in Skopje, Faculty of Computer Science and Engineering, Rudjer Boshkovikj 16, 1000 Skopje, North Macedonia\\
\email{\{miroslav.mirchev,lasko.basnarkov,igor.mishkovski\}@finki.ukim.mk}}
\maketitle              
\begin{abstract}
We investigate searching efficiency of different kinds of random walk on complex networks which rely on local information and one-step memory. For the studied navigation strategies we obtained theoretical and numerical values for the graph mean first passage times as an indicator for the searching efficiency. The experiments with generated and real networks show that biasing based on inverse degree, persistence and local two-hop paths can lead to smaller searching times. Moreover, these biasing approaches can be combined to achieve a more robust random search strategy. Our findings can be applied in the modeling and solution of various real-world problems.
\end{abstract}
\section{Introduction}

Random walk is a ubiquitous concept that describes wandering in certain space in which the location where the walker will be in the next moment is chosen randomly. In complex networks it can applied for modeling diverse phenomena like searching through information networks \cite{Austerweil2012}, diffusion of information, ideas and viruses in social networks, stock market fluctuations, and solving various problems such as page ranking in the web \cite{page1999pagerank}, semi-supervised graph labeling \cite{zhu2003semi,glonek2019semi}, link prediction in graphs \cite{backstrom2011supervised}, and graph representation learning \cite{Grover2016,nikolentzos2020random}.

Since the onset of interest in complex networks, various models of random walk on top of them have been proposed. The standard uniform random walk is based on randomly choosing the next node in the walk with equal probability from all neighbors of the node where the walker currently is. By applying master equation approach \cite{noh2004random} or Markov chain theory \cite{grinstead2012introduction} one can obtain theoretical results for a key quantity in the random walk -- the mean first passage time (MFPT), that represents the expected number of steps needed for the walker to reach randomly chosen target for the first time. Using the same formalism, various modifications of the uniform random walk have been applied that exploit the local properties of the network, aimed at improving the search time. One approach is based on the degrees of the neighbors \cite{fronczak2009biased}, particularly when biasing proportionally to the inverse degree of the next node \cite{bonaventura2014characteristic, basnarkov2020random}. Some authors have considered local neighborhood exploration by random walks using marking as well as biasing based on neighbors degrees\cite{berenbrink2010speeding}. In another approach memory is applied where the probability to jump to some next node depends on the current, but also on the previously visited one \cite{basnarkov2017persistent, basnarkov2020random, cao2021one}. Other problems that have recently received attention are random walk on networks with resetting \cite{riascos2020random}, multiple simultaneous random walks \cite{patel2016hitting}, and random walk on hypergraphs \cite{carletti2020random}.

The theoretical expressions for calculating MFPT in random walks with one-step memory presented in \cite{basnarkov2020random} provide a useful testbed that can be employed for comparing various biasing strategies in relatively small networks. Nevertheless, the findings can be then applied to networks with arbitrary sizes. In this work, we aim to study and combine different approaches with local information in order to see whether further improvement is possible. We study five types of random walks with one-step memory: simple forward going, inverse degree biased, two-hop paths based, persistent, and we introduce a combination of persistent and inverse degree biased. For comparison in our study we also include two standard random walks without memory: uniform and inverse degree biased. Our findings can be applied for potential improvements in the study of a wide range of problems mentioned at the beginning of this introduction.

In Section II we describe the theoretical expressions for calculating MFPTs in random walks with one-step memory on complex networks represented as graphs. Several graph searching strategies using such random walks are described in Section III. In Section IV we present the results obtained with the theoretical expressions and numerical simulations on several synthetic and real complex networks, while in Section V we give some general conclusions.

\section{Mean first passage time of random walks with one-step memory on complex networks}\label{SEC:with_mem}

In this section we briefly restate the main analytical results from \cite{basnarkov2020random} for representing a random walk with a one-step memory over complex networks, but a detailed explanation of the expressions derivation can be found in the original paper. For the sake of simplicity we use notations for undirected networks, although the same theory also holds for directed networks. A complex network given as a graph $G(V, E)$ composed of a set of vertices $V$, $|V|=N$, and a set of edges $E$, $|E|=L$, can be represented by an adjacency matrix $\mathbf{A}_{N \times N}$. We study discrete-time random walks with a one-step memory, so that a random walk that in the previous steps has visited nodes $\{\dots,o,p,q,r\}$ and currently is in node $s$, can visit a next node $t$ with a probability
\begin{equation}
    p(t|s, r, q, p, o, \dots) = p(t|s, r).
\end{equation}
In order to represent such a random walk with a Markov chain instead of using the nodes as states we use the links between the nodes. The transition matrix of the corresponding Markov chain $\mathbf{P}_{L \times L}$ will have elements ${p_{rs, st} = p(t|s, r)}$, ${\forall rs,st, \in E}$. These elements can take arbitrary values that represent probabilities, depending on the chosen random walk.

The random walk can be initialized by starting from node $a$ and then passing to a random neighbor $b$, so from that moment on the transitions can be made according to $\mathbf{P}$. The problem of finding a target node $z$ can be represented as reaching any state $yz$, where $y$ is any neighbor of $z$. This process can be represented by an absorbing Markov chain with a transition matrix $\mathbf{P}_{(z)}$ where all states $yz$ are absorbing, while all other transitions states $ij, j \neq z$ are transient. The theory of absorbing Markov chains and particularly the mean time to absorption (MTA) can be then used to calculate the mean first passage time (MFPT) from $a$ to $z$ \cite{seneta2006non,grinstead2012introduction}. For simplicity we assume that the random walk never starts from the target $z$, so for the MFPT calculation we can safely omit all states $zy,\forall y$.
The transition matrix $\mathbf{P}_{(z)}$ takes the form
\begin{equation}
\label{eq:P_z}
    \mathbf{P}_{(z)} = \begin{vmatrix}
\mathbf{Q}_{(z)}&\mathbf{R}_{(z)}\\
\mathbf{0}&\mathbf{I}\\
\end{vmatrix},
\end{equation}
where $\mathbf{Q}_{(z)}$ is an $(L-k_z) \times (L- k_z)$ matrix containing the transition probabilities among transient states, $\mathbf{R}_{(z)}$ is an $(L -  k_z) \times k_z$ matrix representing the transitions from the transient to the absorbing states, and $\mathbf{I}$ is an $k_z \times k_z$ identity matrix.
The fundamental matrix for the corresponding Markov chain contains the expected number of steps that a random walk starting from any transient state $ab$ is present in another transient state $ij$
can be expressed as the infinite sum

\begin{equation}
    \mathbf{Y}_{(z)} = \mathbf{I} + \mathbf{Q}_{(z)} +\mathbf{Q}_{(z)}^2 + \cdots.
\end{equation}
The powers of $\mathbf{Q}_{(z)}$ diminish as $n \rightarrow \infty$, and $\mathbf{Y}_{(z)}$ converges towards
\begin{equation}
    \mathbf{Y}_{(z)} = \left(\mathbf{I} - \mathbf{Q}_{(z)}\right)^{-1}.\label{eq:Fund_mat_absorbing}
\end{equation}
Then we can obtain a vector containing all the MTA from all possible initial states $ab$ by multiplying with a vector of ones $\mathbf{1}$
\begin{equation}
    \mu_{(z)} = \mathbf{Y}_{(z)}\mathbf{1}.
\end{equation}
Then the MFPT from $a$ to $z$ can be calculated as \cite{basnarkov2020random}
\begin{equation}
    m_{a,z} = 1 + \frac{1}{k_a} \sum_{b \in \mathcal{N}_a} \mu_{(z), ab}.
    \label{eq:MFPT_memory_based_final}
\end{equation}
A Global Mean First Passage Time (GMFPT) \cite{tejedor2009global} can be found by averaging over all starting nodes $a$ as
\begin{equation}
g_z = \frac{1}{N-1} \sum_{\substack{a=1\\a \neq z}}^N m_{a,z}.
\label{eq:GMFPT_def}
\end{equation}
By repeating the same procedure and averaging over all target nodes $z$ we can express a Graph MFPT (GrMFPT) as \cite{bonaventura2014characteristic}
\begin{equation}
    G = \sum_{z=1}^{N} g_z,
    \label{eq:GrMFPT}
\end{equation}
which will be used in the rest of the paper for comparing several different strategies of random walks with one-step memory.

\section{Graph search algorithms based on random walks} \label{sec:algorithm}

In this section we describe several different strategies for graph search using random walks with memory. We also include two classical random walks without memory: a uniform random walk and an inverse degree biased random walk. In a previous work \cite{basnarkov2020random}, we considered the application for a random walk searching strategy based on the number of two-hop paths towards the next node in the walk, which we call "two-hop random walk with memory". However, the results showed that in directed complex networks this strategy does not brings improvements and simply choosing the next node solely based on its inverse degree resulted in shorter hitting times. Therefore, in this paper we also consider four other random walk strategies with one-step memory. The first one simply avoids going back, which was thoroughly studied in \cite{cao2021one}, and we refer it as "forward random walk with memory". The second strategy, which we call "inverse degree random walk with memory", in addition to avoiding going backwards chooses the next node based on its degree. Another approach called "persistent random walk with memory" which employs biasing towards more distant nodes by avoiding neighbors of the previously visited node, was numerically studied in \cite{basnarkov2017persistent}, but here we further provide calculations based on the theoretical expressions. Moreover, we examine a hybrid of the persistent and the inverse degree random walks with memory, in order to combine their strengths and help cover their weaknesses. For calculating MFPTs and GrMFPT in the random walks with memory we will use the theoretical expressions from \cite{basnarkov2020random}, described in the previous section, for all networks in our study. However, in the random walks without memory we will use the expressions presented in \cite{basnarkov2020random} for finding GrMFPT in the generated networks, and standard expressions based on absorbing Markov chains \cite{seneta2006non} for the real networks due to numerical problems with the expressions from \cite{basnarkov2020random}.

\subsection{Classical random walks without memory}

\subsubsection{Uniform random walk (U-RW)} - at each step the random walk makes a transition from node $s$ to any of its neighbors $t$ with an equal probability ${p_{st}=1/k_s}$.

\subsubsection{Inverse degree random walk (ID-RW)} - the visiting probability of $s$ to a neighboring node $t$ is inversely proportional to its node degree $1/k_t$, hence
\begin{equation}
p_{st}=\frac{1/k_t}{\sum_{t\in \mathcal{N}_s} 1/k_t}.
\end{equation}

\subsection{Random walks with memory}

In the random walks with memory, we denote the previous visited node as $r$, the current node as $s$, and the potential next nodes as $t$.

\subsubsection{Forward random walk with memory (F-RWM)} - the random walk avoids going back, by exploiting the one-step memory, unless there is no way to keep going forward. The probability of visiting the other neighboring nodes is equal and expressed as
\begin{equation}
p_{rs,st}=1/(k_s-1), \forall r \neq t,
\end{equation}
while the probability of going back can be written as
\begin{equation}
  p_{rs,sr} =
    \begin{cases}
      1 & \text{if r is the only neighbor of s,}\\
      0 & \text{otherwise.}
    \end{cases} 
    \label{eq:forward}
\end{equation}

\subsubsection{Inverse degree random walk with memory (ID-RWM)} - if a random walk has previously visited $r$ and currently is in $s$ would visit some of its other neighbors with a probability.
\begin{equation}
p_{rs,st}=\frac{1/k_t}{\sum_{t\in \mathcal{N}_s\setminus \{r\}} 1/k_t},
\end{equation}
while it avoids going back to $r$ by following Eq. (\ref{eq:forward}).

\subsubsection{Two-hops random walk with memory (2H-RWM)} - the probabilities are proportional to the number of two-hop paths that lead toward the target nodes, so the visiting probabilities are given as
\begin{equation}
p_{rs,st} = \frac{\frac{1}{b_{rt}}}{\sum_{u\in \mathcal{N}_s}\frac{1}{b_{ru}}}, 
\end{equation}
where $b_{rt}$ is the number of two-hop paths between $r$ and $t$, or the elements of $\mathbf{B} = \mathbf{A}^2$.

\subsubsection{Persistent random walk with memory (P-RWM)} - the random walk avoids going backwards or towards the neighbours of the previously visited node. Let us denote with $N_1=|\mathcal{N}_r \cap \mathcal{N}_s|$ the number of common neighbors of $s$ with $r$ which it visits each with a probability $p_1$, and let $N_2=|\mathcal{N}_s \setminus \{\mathcal{N}_r \cap \mathcal{N}_s\}|$ be the number of other neighbors, which it visits with a probability $p_2$ each, and let $p_0$ be a probability of immediately going back to $r$. We can then write $p_0 + N_1 p_1 + N_2 p_2 = 1$. By introducing the parameters $p_2 / p_1 = \alpha$ and $p_0 / p_1 = \beta$, the previous expression can be rewritten as $p_1 \left(\beta + N_1 + \alpha N_2 \right) = 1$. Hence, the probability of going back to $r$ is given by

\begin{equation}
  p_{rs,st} =
    \begin{cases}
      p_0=\frac{\beta}{C}   & \text{if } r=t,\\
      p_1=\frac{1}{C}       & \text{if } t \in \mathcal{N}_r \cap \mathcal{N}_s,\\
      p_2=\frac{\alpha}{C}  & \text{if } t \in \mathcal{N}_s \setminus \{\mathcal{N}_r \cap \mathcal{N}_s\},
    \end{cases} 
\end{equation}
where $C=\beta + N_1+\alpha N_2$ is a normalization coefficient. A detailed analysis of the effects of the parameters $\alpha$ and $\beta$ on GrMFPT can be found in \cite{basnarkov2017persistent}.

\subsubsection{Persistent inverse degree random walk with memory (PID-RWM)} - combines the strengths of the persistent and inverse degree biased random walks. We have excluded the possibility for going back to $r$ so $p_{rs,sr}=0$, except when it is the only option $p_{rs,sr}=1$. The probability of going toward a common neighbor with $r$ will become

\begin{equation}
  p_{rs,st} =
    \begin{cases}
      \frac{1/k_t}{C}       & \text{if } t \neq r, t \in \mathcal{N}_r \cap \mathcal{N}_s,\\
      \frac{\alpha/k_t}{C}  & \text{if } t \neq r, t \in \mathcal{N}_s \setminus \{\mathcal{N}_r \cap \mathcal{N}_s\},
    \end{cases} 
\end{equation}
where the normalization coefficient is 
\begin{equation}
C=\sum_{t \in \mathcal{N}_r \cap \mathcal{N}_s}1/k_t + \sum_{t \in \mathcal{N}_s \setminus \{\mathcal{N}_r \cap \mathcal{N}_s\}} \alpha/k_t.
\end{equation}
Once again the random walk avoids going back to $r$ by following Eq. (\ref{eq:forward}). We like to note that for $\alpha=1$, PID-RWM becomes identical to ID-RWM.

\section{Results}

\begin{figure}[!tb]
    \subfloat[BA \label{fig:GMFPTBANet}]{
        \includegraphics[width=0.49\columnwidth]{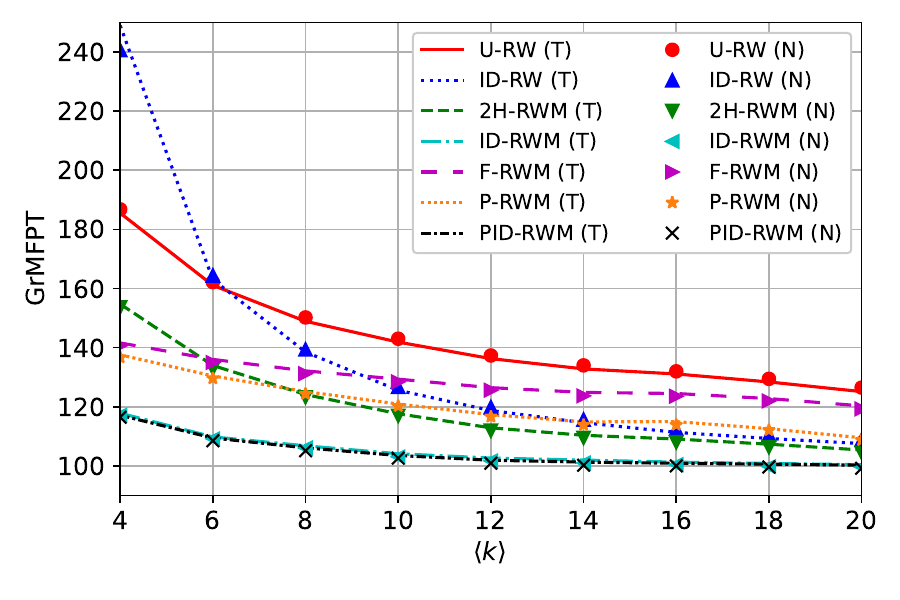}
    }
    \subfloat[WS \label{fig:GMFPTWSNet}]{
    	\includegraphics[width=0.49\columnwidth]{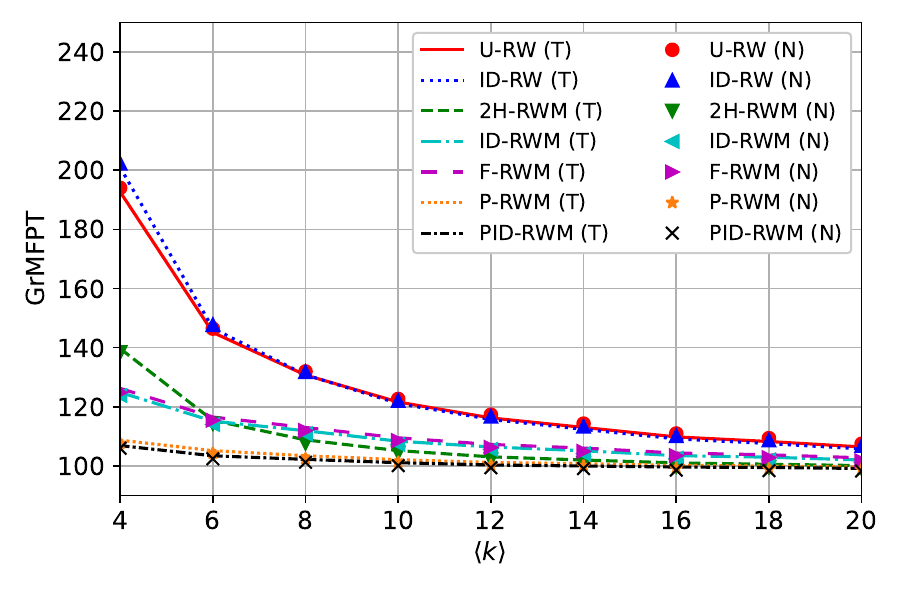}
    }\\
    \subfloat[ER \label{fig:GMFPTERNet}]{
    	\includegraphics[width=0.49\columnwidth]{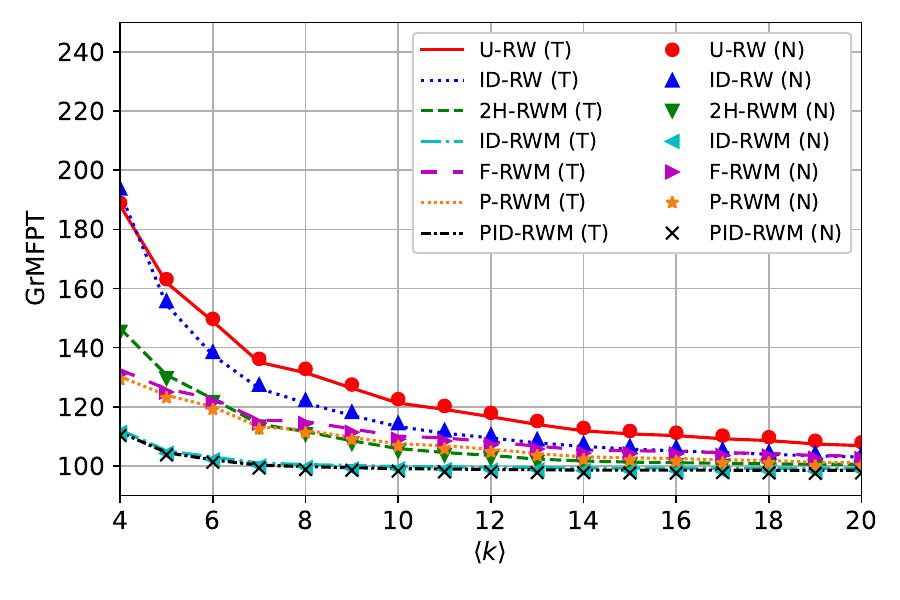}
    }
    \subfloat[ER directed \label{fig:GMFPTERDirect}]{
		\includegraphics[width=0.49\columnwidth]{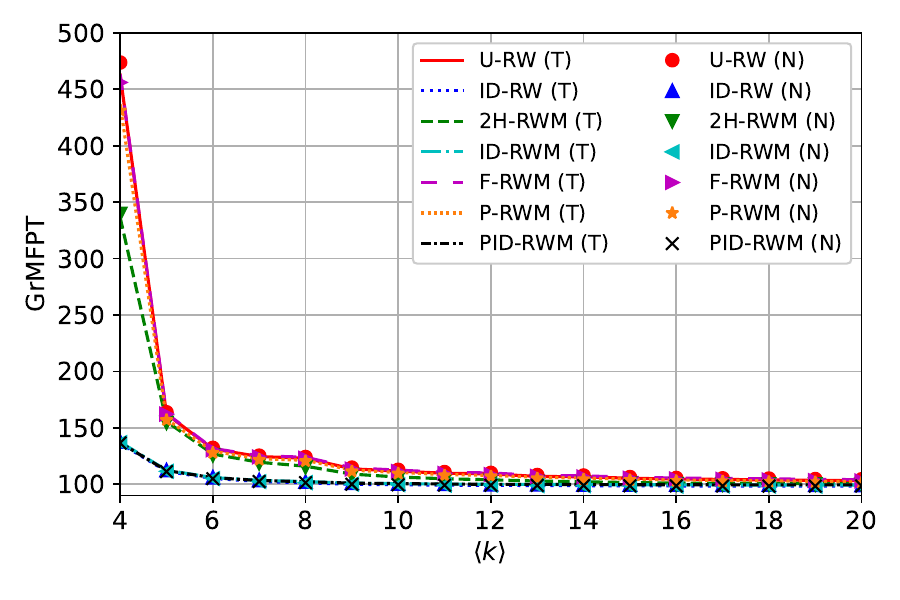}
	}
    \caption{GrMFPT in (a) BA, (b) WS, (c) ER, and (d) ER directed networks, with $100$ nodes and varied average node degree $\langle k \rangle$ for $7$ different random walks. The lines are theoretical values (T) and the markers numerical estimates (N).}
    \label{fig:GMFPTNet}
\end{figure}

First, we examine the GrMFPT for five types of random walks: classical, inverse degree, two-hop with memory, inverse degree with memory and forward with memory, for three complex networks models: Barab{\'a}si-Albert (BA), Watts-Strogatz (WS), and Erdős–Rényi (ER) with undirected and directed links. In Figure \ref{fig:GMFPTNet} we show the results calculated using the theoretical expressions and by numerical simulations of the random walks transitions. The results are averaged over $10$ different network instances generated with the same parameter values, while the numerical simulations are further averaged across $10$ repetitions of all node pairs. The rewiring probability in the WS model is $p_\mathrm{rew}=0.2$. In P-RWM and PID-WM, $\alpha=10$ and $\beta=0.01$, but one can further analyse the effects of varying $\alpha$. As can be seen in Figure \ref{fig:GMFPTBANet}, for BA networks the ID-RWM and PID-RWM significantly outperform the other methods, and the difference is large for small $\langle k \rangle$ particularly with the similar ID-RW, which also employs inverse degree biasing but lacks memory. The results for WS networks presented in Figure \ref{fig:GMFPTWSNet} show how the persistent random walks exploit the memory, with PID-RWM slightly outperforming P-RWM. In ER networks (Figure \ref{fig:GMFPTERNet}), the inverse degree biasing proves crucial again and the ID-RWM and PID-RWM show best searching performance. As we have previously noted in \cite{basnarkov2020random}, in directed ER networks (Figure \ref{fig:GMFPTERDirect}) the simple inverse degree biasing without memory, still holds well with almost identical performance with the other inverse degree biased approaches ID-RWM and PID-RWM. These results can be expected as in sparse ER directed networks rarely ever two nodes are connected in both directions, hence, the going back avoidance is rarely exploited. In some real networks these bidirectional mutual connectivity can be present more often so the memory could be more beneficial there. Overall, we can conclude that the PID-RWM shows the best and most consistent performance across different network topologies and densities. 

\begin{figure}[!tb]
	\subfloat[BA \label{fig:KL_BA}]{
		\includegraphics[width=0.49\columnwidth]{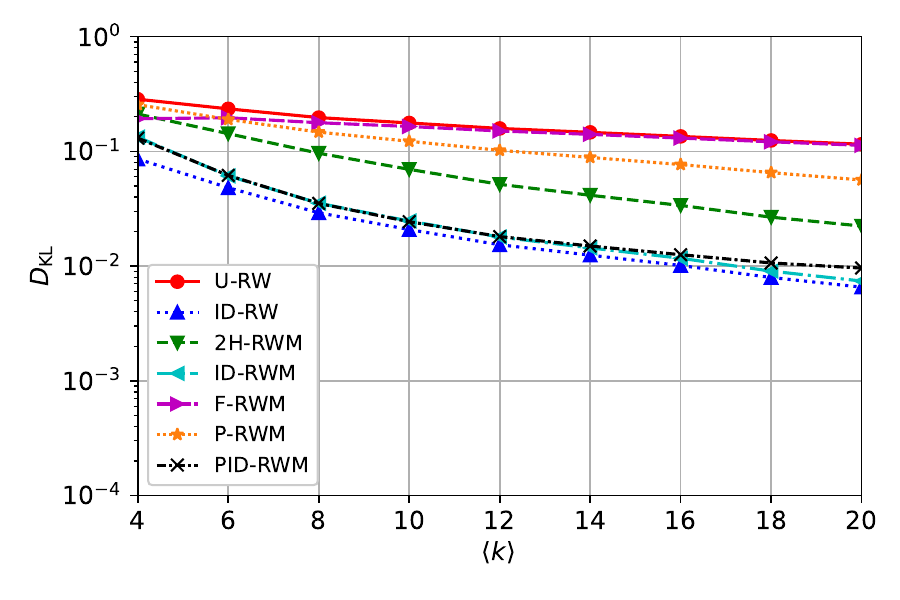}
	}
    \subfloat[WS \label{fig:KL_WS}]{
		\includegraphics[width=0.49\columnwidth]{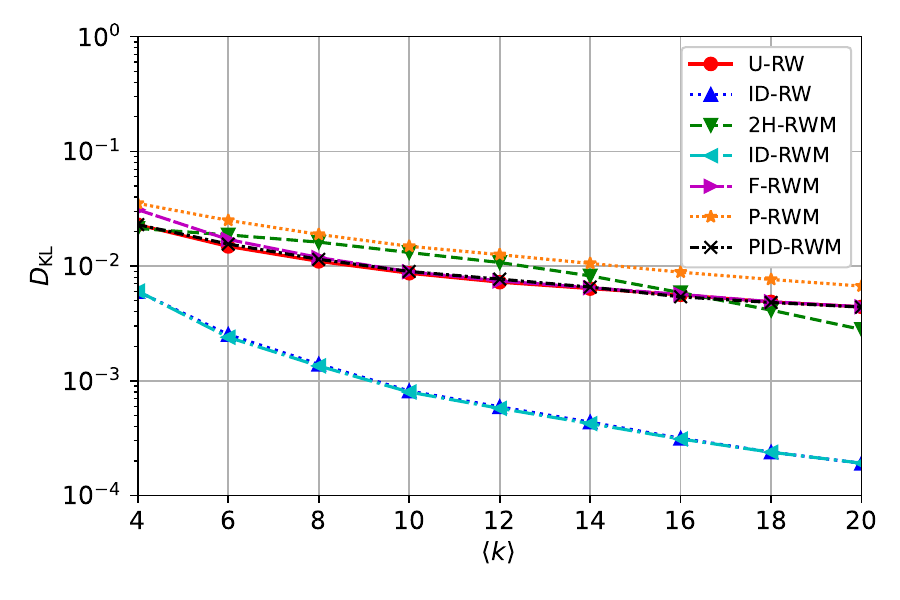}
	}\\
	\subfloat[ER \label{fig:KL_ER}]{
		\includegraphics[width=0.49\columnwidth]{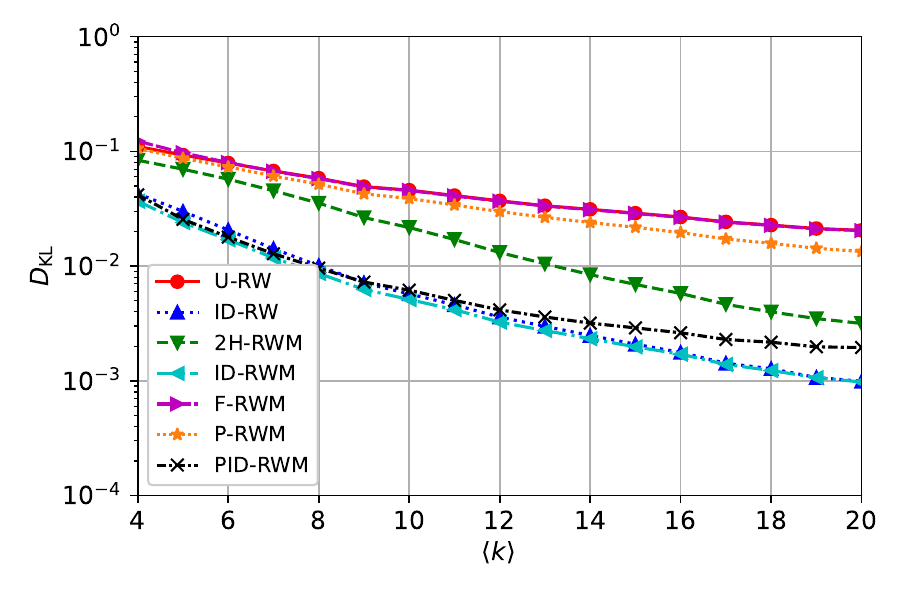}
	}
    \subfloat[ER directed \label{fig:KL_ER_directed}]{
		\includegraphics[width=0.49\columnwidth]{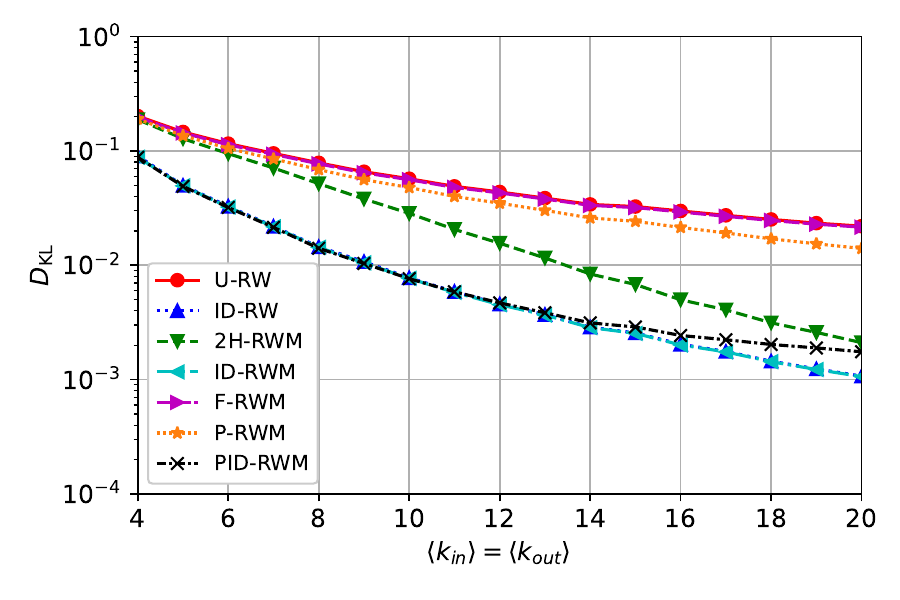}
	}
\caption{Kullback-Leibler divergence of the stationary occupation probability of $7$ different random walks from a uniform density in (a) BA,  (b) WS, (c) ER, and (d) ER directed networks, with $100$ nodes and varied average node degree $\langle k \rangle$.
\label{fig:KL}}
\end{figure}

We have also studied how the stationary distributions of the visiting probabilities are affected from the choice of the random walk strategy, by comparing them with a uniform distribution using the KL divergence

\begin{equation}
    D_{\textrm{KL}}(P||Q) = \sum_{i} P(i) \log \frac{P(i)}{Q(i)},
\end{equation}
and the results are shown in Figure \ref{fig:KL}. As expected the ID-RW significantly equalizes the visiting probabilities, which is the reason behind the often observed shorter search times compared to the U-RW. In most cases ID-RW achives lowest $D_{KL}$, however, for ER networks with low $\langle k \rangle$ ID-RWM and PID-RWM achive slightly lower $D_{KL}$.
  
\begin{figure}[!tb]
	\subfloat[WS model with $k=4$ \label{fig:WS_K=4_varProb}]{
		\includegraphics[width=0.49\columnwidth]{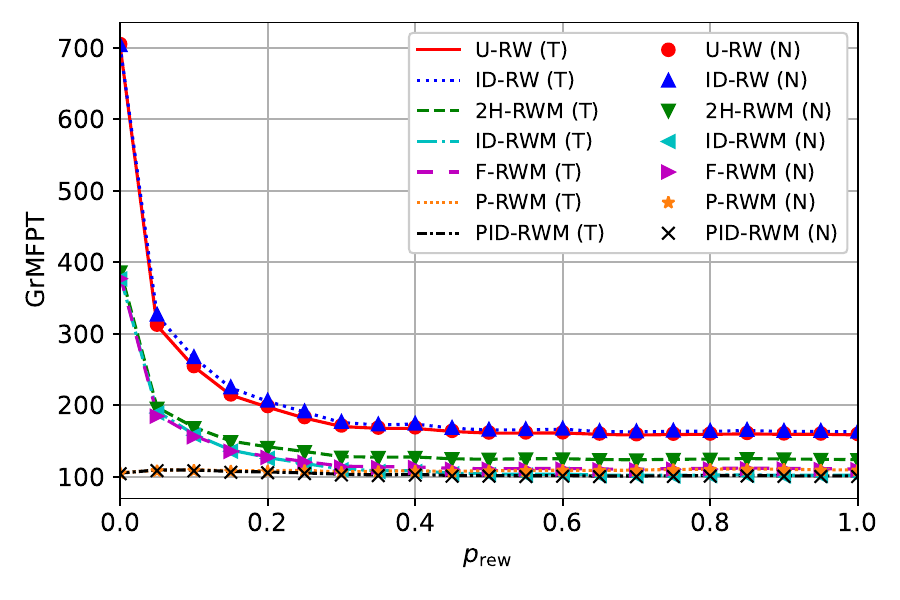}
	}
	\subfloat[WS model with $k=6$ \label{fig:WS_K=6_varProb}]{
		\includegraphics[width=0.49\columnwidth]{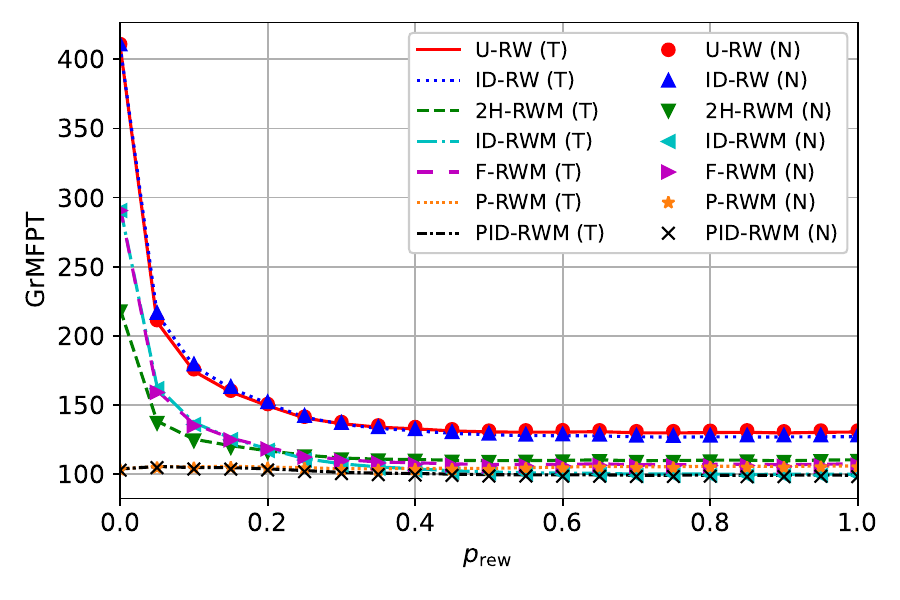}
	}

\caption{GrMFPT in WS networks with (a) $k=4$, and (b) $k=6$, composed of $100$ nodes with varied rewiring probability for seven different random walks. The lines are theoretical values (T) and the markers numerical estimates (N).
\label{fig:WS_varProb}}
\end{figure}

To gain a deeper insight into the effect of the rewired links in the WS model, we studied how the MFPT varies with the change of the rewiring probability $p_\mathrm{rew}$ in networks with $k=4$ and $k=6$ neighbors per node, and the results are given in Figure \ref{fig:WS_varProb}. The WS model transitions from a regular lattice toward a completely random ER network, as $p_\mathrm{rew}$ is varied from 0 to 1. It can be seen how for small $p_\mathrm{rew}$ PID-RWM and P-RWM behave similarly and have best performances, however, as $p_\mathrm{rew}$ increases the performance of P-RWM decreases eventually loosing the pace with ID-RWM, while PID-RWM keeps its performance at level with ID-RWM.

\begin{table}[h]
\centering
		\caption{Statistical properties of the real networks: number of nodes $N$, number of links $L$, density $D$, average node degree $\langle k \rangle$, average clustering coefficient $C$, average path length $\langle l \rangle$,  and diameter $d$.\newline}
		\label{tab:properties}
		\begin{tabular}{c|c|c|c|c|c|c|c|c}
			\hline
			      	      &Type         &$N$  &$L$    &$D$     &$\langle k \rangle$   & $C$    & $\langle l \rangle$  &$d$\\
			\hline
    		Internet        &undirected   &6474 &13233  &0.0006  &4.29	            &0.2522  &3.7050                &9\\
			Wikipedia 	    &directed     &4051 &119000 &0.0068  &27.62	            &0.1892  &3.1813                &9\\
            Euroroad        &undirected   &1039 &1305   &0.0024  &2.51              &0.01890 &18.3951	            &62\\
            FB-Pages        &undirected   &620  &2102   &0.0109  &3.39              &0.3309  &5.0887                &17\\
            Bio-diseasome   &undirected   &516  &1188   &0.0089  &2.30              &0.6358  &6.501                 &15\\
            CA-netscience   &undirected   &379  &914    &0.0128  &2.41              &0.7412  &6.042                 &17
		\end{tabular}
\end{table}

\begin{figure}[!tb]
\centering
	\subfloat[Euroroad network \label{fig:Euroroad}]{
		\includegraphics[width=0.24\columnwidth]{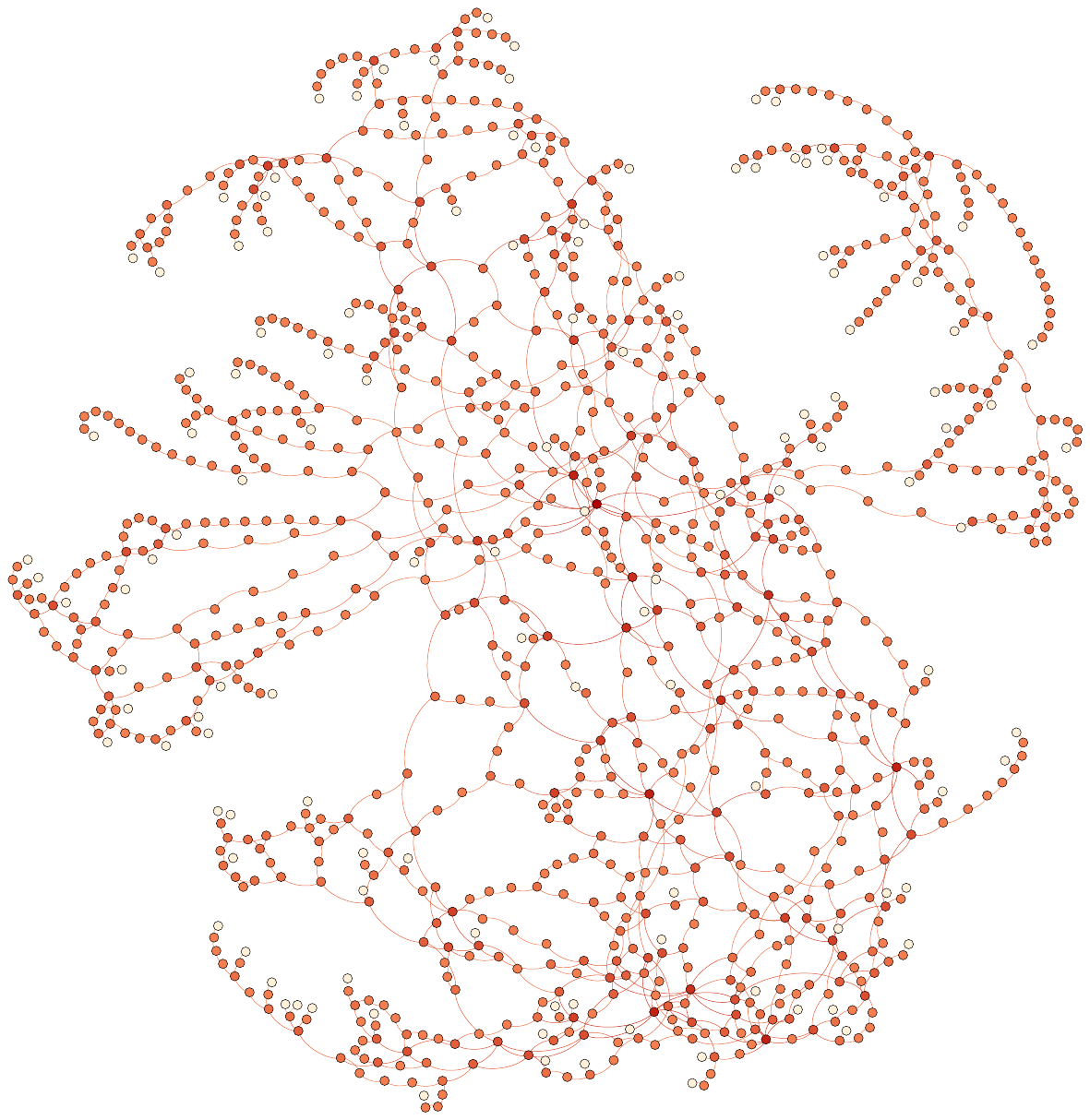}
	}	
    \subfloat[FB-Pages \label{fig:Internet}]{
		\includegraphics[width=0.24\columnwidth]{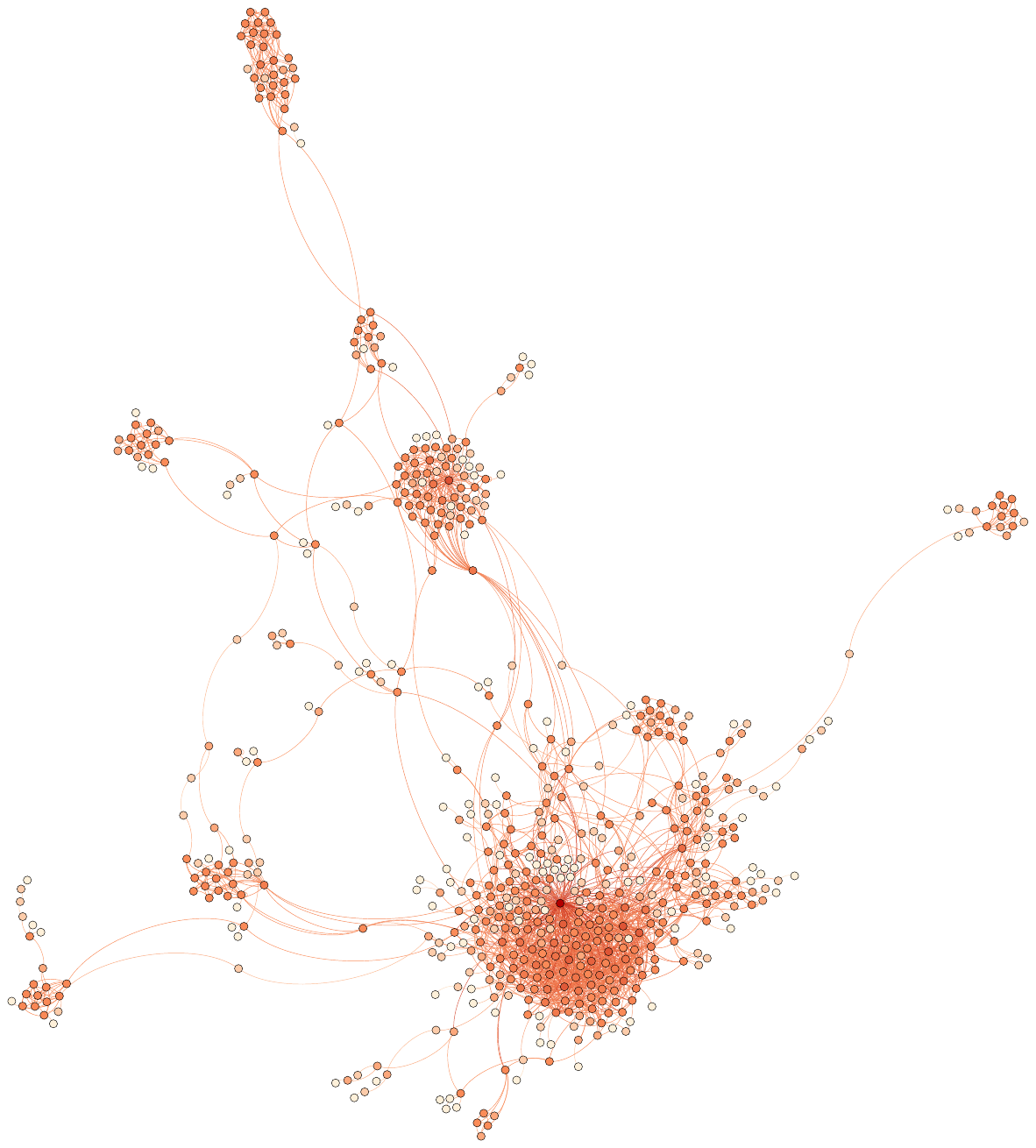}
	}
	\subfloat[Bio-diseasome \label{fig:Internet}]{
		\includegraphics[width=0.24\columnwidth]{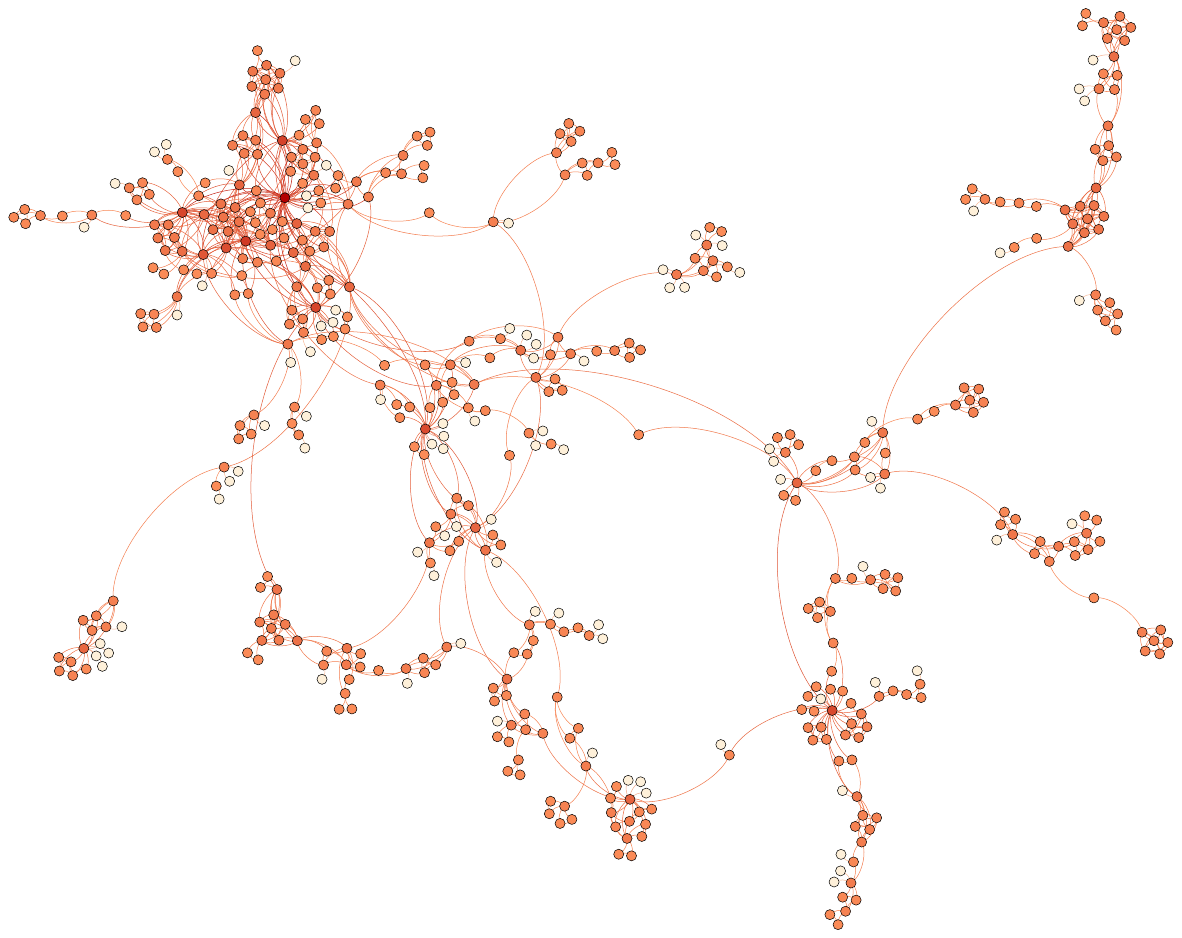}
	}
    \subfloat[CA-netscience \label{fig:Internet}]{
		\includegraphics[width=0.24\columnwidth]{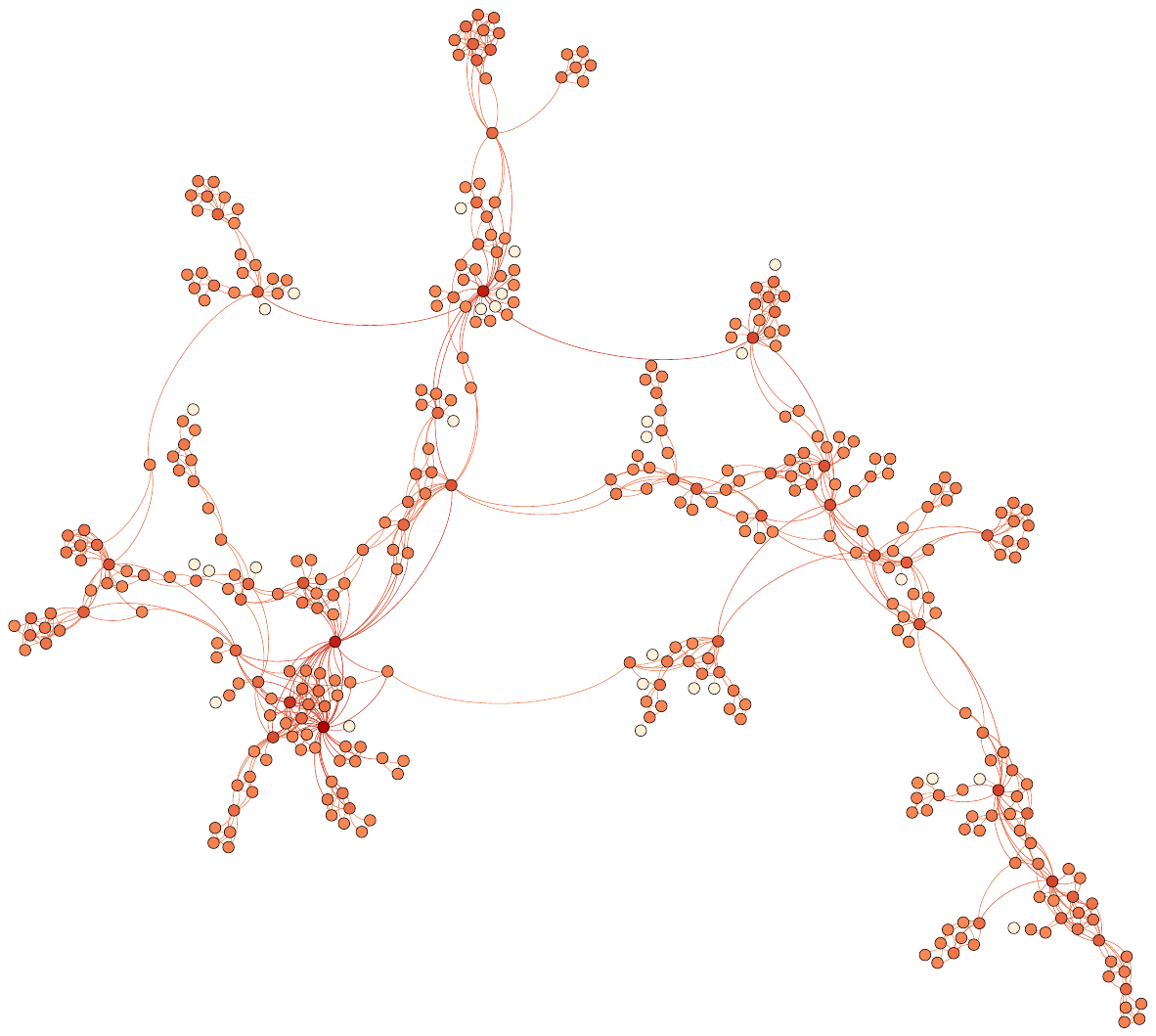}
	}
\caption{Visualization in Gephi of four real networks topologies, where a darker color indicates a larger node degree.
\label{fig:realnetworks}}
\end{figure}

We also made comparison of the various random walks on several real networks and their main structural properties are given in Table \ref{tab:properties}. The first network is a representation of the Internet at level of autonomous systems derived from BGP logs \cite{leskovec2005graphs}, which is known to have the scale-free property. The second network is an excerpt from Wikipedia pages \cite{west2009wikispeedia,west2012human}, also having a scale-free property. This network was used in \cite{west2012human} to study human wayfinding to a given target through Wikipedia pages. The original dataset consists of $4592$ nodes and $119882$ links, but for our analysis we use the largest strongly connected component. The third network Euroroad is a representation of major European roads \cite{vsubelj2011robust}. It is an undirected network and consists of $1174$ nodes and $1417$ edges, from which we take the largest connected component. The fourth network FB-Pages is a collection of Facebook pages and their mutual likes \cite{rozemberczki2019gemsec}, and the fifth is a network of human diseases (Bio-diseasome) \cite{goh2007human}. The sixth network CA-netscience depicts collaboration in publications between researchers in the field of network science \cite{newman2006finding}. The first two datasets are taken from the SNAP dataset collection \cite{leskovec2016}, and the last three from the Network repository \cite{rossi2015}. A visualization of the last four networks is given in Figure \ref{fig:realnetworks}, using the Force atlas layout, where nodes with larger degree are colored darker.

\begin{table}[h]
		\caption{GrMFPT for six real networks with various random walks, where (T) indicates results with theoretical expressions and (N) with numerical simulations.\newline}
		\label{tab:1}
		\begin{tabular}{l|r|r|r|r|r|r|r}
			\hline
			      	 &U-RW    & ID-RW   & 2H-RWM  & ID-RWM  & F-RWM  & P-RWM & PID-RWM\\

			\hline
            Internet (N)    &19385	&178916	&18293	&23410	&17318	&16443	&20260\\ 
            \hline
            Wikipedia (N)   &22974882	&11342	&1269384   &10930	&29894802	&9086546	&11857\\
            \hline
            Euroroad (N)        & 9246	&12854	&5489	&2968	&2742	&2714	&2900\\
            Euroroad (T)        &9243   &12762  &5485	&2954	&2760	&2716	&2922\\
            \hline
            FB-Pages (N) &3516	    &3879	  &1673	    &1588	&2422	&1845	&1450\\
            FB-Pages (T) &3521	    &3855	&1676	&1568	&2401	&1846	&1453\\

            \hline
            Bio-diseasome (N) &3526	   &8536	 &2114	&4277	&2471	&1663	&3246\\
            Bio-diseasome (T) &3488	   &8503     &2119	&4322	&2474	&1662	&3300\\

            \hline
            CA-netscience (N) &1895	&4747   &1287	&2326	&1409	&1046  &2488\\
            CA-netscience (T) &1891 &4742   &1297  &2354  &1409  &1062  &2694\\
		\end{tabular}
\end{table}

The results with the real networks summarized in Table \ref{tab:1} show that the theoretical expressions are in accord with the numerical simulations. The first two networks are relatively larger, so for them we conducted only numerical simulations, while for the other networks we also provide results from the theoretical expressions. The numerical results for all networks are obtained by calculating the MFPT between $100000$ randomly chosen node pairs. There were some numerical problems in the calculations of the GrMFPT for some networks using the analytical expressions for memoryless random walks from \cite{basnarkov2020random}, so for them we used standard expressions based on absorbing Markov chains \cite{seneta2006non}. The results indicate that the random walks can behave differently in real networks, as their structure is not always very similar to networks generated with classical models. For example, P-RWM is better than PID-RWM and ID-RWM, which was often not the case in the generated networks. P-RWM is better for most networks, but for Wikipedia it is significantly worse than ID-RWM and PID-RWM and it also fails behind them for FB-pages. Another interesting observation is that from all considered real networks only Wikipedia is directed and has a very high average node degree, hence, the inverse degree biased group of random walks show best results. In this network the forward going behavior is naturally enforced, while the inverse degree biasing flattens the visiting probabilities and speeds up the search.

\section{Conclusion}\label{SEC:conclusions}

We studied various types of graph searching algorithms based on biased random walks using local information and a one-hop memory, which can be applied in modelling real-world phenomena and solving various problems. The results calculated with the given theoretical expressions match those with the numerical simulations both for generated and real networks. Generally biasing can be helpful, particularly in undirected networks, however, it should be applied carefully as different strategies could produce varying results depending on the specific network properties. For example, biasing based on inverse degree can be useful in networks with a scale-free property, but it can be unfavourable in networks with large transitivity. Moreover, the application in real networks can lead to slightly different results from what is obtained in supposedly similar generated networks. As a future work one can expand the study and include multiple random walkers, however, in this way the transition and state matrices would increase exponentially with the number of walkers. Another possible direction is considering memory in random walk with restart or teleportation, or random walk on hypergraphs.

\section*{Acknowledgement}
This research was partially supported by the Faculty of Computer Science and Engineering, at the Ss. Cyril and Methodius University in Skopje, N. Macedonia. 

\bibliographystyle{splncs04}
\bibliography{Paper1668}

\end{document}